\documentclass{article}
\usepackage{graphicx}        
\usepackage{color}           
\usepackage{hyperref}
\usepackage{breakurl}        
\usepackage{natbib}

\newcommand{\etal}{{\it et al.}}

\begin{document}
\sloppy

\title{ Periodic change of polarity of photospheric magnetic field }
\author{E.~S.~Vernova; M.~I.~Tyasto\\
IZMIRAN SPb Filial, St. Petersburg, Russia\\
elenavernova96@gmail.com\\
D.~G.~Baranov\\
Ioffe Institute, St. Petersburg, Russia
}
\date{}
\maketitle
\begin{abstract}
The work is devoted to the study of periodic structures in the magnetic field of the photosphere. The polarity
 of the Sun's magnetic fields shows cyclicity with periods of 1--3 years, which is possibly due to quasi-biennial
 variations that are found in various parameters of solar activity, interplanetary space, and in cosmic rays. In this work,
variations in weak magnetic fields of the photosphere were investigated, which required the use of special data processing
methods. Synoptic maps of the photospheric magnetic field for the period 1978--2016 (NSO Kitt Peak) were used. To highlight
the contribution of weak magnetic fields, the saturation threshold for synoptic maps was set at 5\,G. Based on thus transformed
synoptic maps, a time-latitude diagram was constructed. For the selected latitude intervals, smoothing, trend removal and 
Fast Fourier Transform (FFT) were performed. In the time-latitude diagram, 6 spatiotemporal
regions with distinct cyclic structure were observed during 4 solar cycles (21--24). The periods ranged from 0.5 to 4 years
with a maximum at 1.2 years. The amplitude of ripples is significantly higher for those intervals in which the polar field
had positive sign. This effect confirms the connection of ripples variations with the polarity of the 22-year magnetic cycle.
\end{abstract}

\section{Introduction}
The Sun's magnetic field, its distribution over the Sun's surface, and its evolution over time are responsible for all
manifestations of solar activity (SA). Groups of magnetic fields of various magnitudes, from the strongest magnetic
fields to background magnetic fields, are associated with certain solar phenomena. Cyclic changes in solar activity
reflect corresponding  periodic changes in the Sun's magnetic field. Magnetic fields follow a 22-year magnetic cycle (Hale cycle): the
law of polarity reversal, which manifests itself as a change in the  sign of the polar field near the SA maximum and
as a change in the sign of the leading and trailing sunspots near the SA minimum. The change in SA with an 11-year cycle is
manifested in a change in the latitudinal distribution of SA in the form of the so-called Maunder butterflies. When
studying the 11-year cycle of solar activity, one mainly considers the processes associated with sunspots, that is, with the strongest
magnetic fields.

The patterns of distribution of weak magnetic fields are not as well understood as the distribution of strong
magnetic fields, although it is the weak fields that occupy most of the Sun's surface. According to our estimates
\citep{vern1} based on NSO Kitt Peak data for the period 1978--2016, magnetic fields with strength
$| B |\leq 10$\,G occupied more than 82\% of the surface area of  the Sun. The space-time evolution of weak magnetic
fields has been discussed in \citep{geta1,geta2,murs}.
An important feature of magnetic field cycles is the rearrangement of the distribution of magnetic fields of different
polarities over the surface of the Sun. This rearrangement reflects the processes taking place not only at the visible
surface of the Sun, but also in deeper layers, under the photosphere.
The process of evolution of the solar polar field is greatly influenced by the transfer of magnetic fields by field flows
in the photosphere (surges), which can also be seen in the time-latitude diagram. Here, Rush-to-the-Poles (RTTP) flows play
a special role, which are directly related to the inversion of the polar field. This phenomenon was studied in the green coronal
line \citep{alt}, numerous episodes of RTTP were found in the work \citep{gopa} in occurrence of high-latitude 
prominence eruptions. The role of RTTP flows in the transfer of photospheric magnetic fields was investigated in \citep{petr,mord}.
These flows, drifting from latitudes $\sim 40{^\circ}$ to the poles, are the product of the decay of trailing
sunspots. The lifetime of ``Rush-to-the-Poles'' always falls on the era of maximum solar activity. Their arrival to the poles
of the Sun causes a change in the sign of the polar field and coincides with the polarity reversal of the polar field.

A new phenomenon in the distribution of magnetic fields, consisting in the appearance of wave structures with a period of about
2 years migrating to the poles, was found in {\citep{vecc,ulri}. Using NSO Kitt Peak data processed
by Empirical Mode Decomposition, \citep{vecc} found migration of magnetic fields to the poles during the maximum and
decay of the solar cycle, which they associated with the manifestation of quasi-biennial variations. This phenomenon was
investigated in detail in \citep{ulri}, where the term ``ripples'' was proposed for such magnetic fluxes. In \citep{ulri}
 ripples were discovered using time-latitude diagram differentiation (Mount Wilson Observatory data). One more
method of processing of the time-latitude diagram (deviation from the trend) also showed the alternation of ripples
with opposite polarities. It should be noted that for each of the above methods, ripples were observed at any level of solar
activity. Thus, despite the use of various data and methods for their processing, a number of similar results was obtained
\citep{vecc,ulri} indicating the occurrence at low latitudes and drift to the poles of wave structures with a period close
to the period of quasi-biennial variations.

In our work \citep{vern2}, to study the phenomenon of ripples in weak magnetic fields on synoptic maps of the
photospheric magnetic field, only weak fields were left before plotting time-latitude diagrams. Level of saturation was set
at magnetic field values  of $\pm 5$\,G. As a result, ripple flows were observed that existed between two consecutive RTTPs. The
lifetime of ripples includes decrease, minimum and increase  phases of the solar cycle. These flows of alternating
polarity (each polarity is preserved for 0.5--1 year) drift from the equator to latitudes $\sim 50^\circ$. Ripple flows are
generated by weak fields less than 15\,G, in contrast to RTTP, which are produced by fields of order 100\,G and higher. Also, RTTP
flows always have polarity that coincides with the polarity of the trailing sunspots. They exist for about three years, appearing
at latitudes around $30^\circ$ near the maximum of solar activity.  Then they drift towards the poles, causing polar field reversal.
These significant differences in the properties of ripples and RTTP indicate a different origin of these phenomena.
This work continues the study of the phenomenon of ripples in the weak magnetic fields of the Sun which we started in
\citep{vern2}. Its purpose is to consider the periodic structure of ripples change for solar cycles 21--24.

\section{Data and Method}

The initial data used in the work are synoptic maps produced by the National Solar Observatory Kitt Peak (NSO Kitt Peak) based on
measurements of the Sun's magnetic field made by the Kitt Peak Vacuum Telescope (KPVT) in 1976--2003
(ftp://nispdata.nso.edu/kpvt/synoptic/mag/) and by the equipment Synoptic Optical Long-term Investigations of the Sun (SOLIS)
in 2003--2016 (https://magmap.nso.edu/solis/archive.html). Before 1978 the number of missing field data was very large, that is
why we  used for our analysis data for 1978--2016 only,  making up an array of 521 synoptic maps each representing one Carrington
rotation.  Each synoptic map is a 180 by 360 pixel matrix containing magnetic field strength values in gauss with a longitude
resolution of $1^\circ$ and 180 steps in the sine of latitude. Due to the inclination of the axis of rotation of the Sun, part of the
time it becomes impossible to observe the circumpolar regions. Filling gaps in the data by extrapolating measurements gives less
reliable results. The source of error when studying the magnetic fields of the Sun using synoptic maps is random fluctuations
(noise), which in the polar regions is about 2\,G according to estimates of \citep{harv}. However, averaging 360 values   of
the field over longitude when plotting a time-latitude diagram reduces the error due to random factors to a value of about 0.1\,G.

Synoptic maps averaged over longitude, taking into account the sign of the magnetic field, were used in the construction of the
time-latitude diagram. Since we were primarily interested in weak fields, we limited the influence of strong fields, thereby
emphasizing the contribution of weak fields to the picture of the distribution of the magnetic field over the surface of the Sun.
For this purpose, a saturation threshold of 5\,G was set for each synoptic map. As a result, only
fields with $| B |\leq 5$\,G were left unchanged on each synoptic map, while larger or smaller fields were replaced by the corresponding
limit values   of $+5$\,G or $-5$\,G. The maps thus transformed were used to plot a time-latitude diagram.
In this diagram, some fixed latitudes were selected and temporal changes in magnetic fields at these latitudes were analysed.
To isolate long-period variations, the primary data was processed taking a moving average over 21 Carrington rotations.
Then the deviation from the trend was determined, which was then averaged again by 5  Carrington rotations. This procedure
is illustrated by Figure~\ref{method}, which shows the change of the magnetic field along the latitude $- 33^\circ$ (southern hemisphere) for the
1980--1990 time interval. Figure~\ref{method}a shows the primary values as well as the trend for 21 rotation. Figure~\ref{method}b shows the deviation
of the primary data from the trend. In figure~\ref{method}c, deviations from the trend were smoothed by 5 rotations. The resulting data
were used to investigate periodic changes in ripples.
\begin{figure}[h]
  
	\centerline{\includegraphics[width=0.95\textwidth,clip=]{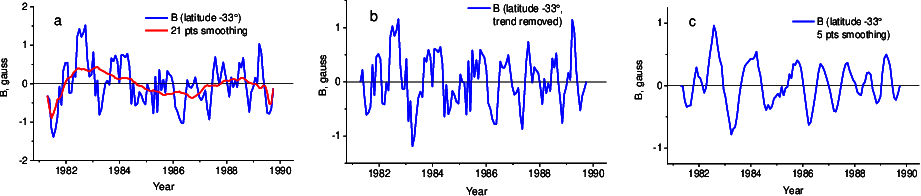}
              }  
\caption{  Removing trend from magnetic field data. (a) Change of magnetic field $| B |\leq 5$\,G along latitude $-33^\circ$,  in the interval S1 (1982--1989).
The trend was determined as a result of taking a moving average over 21 points.
(b) Magnetic field after trend removal.
(c) Magnetic field (b) averaged over 5 points.
}

      \label{method}                                        
   \end{figure}

\section{Results and Discussion}

Studies of \citet{vecc} and \citet{ulri} show that ripples change with a period of about two years. In \citep{vecc} these variations are considered as one of the manifestations of quasi-biennial oscillations (QBOs) during the period of maximum and decay of the solar cycle. A  period of 0.8 to 2 years was found for ripples in \citep{ulri}. In our work \citep{vern1} the averaged values of the period of ripples for the interval of latitudes from the equator to $50^\circ$  were obtained by two methods: a) using the method of empirical orthogonal functions (EOF) and b) summing the time profiles of the magnetic field taking into account their shift in latitude over time. The periods obtained by these two methods were 1.8 and 1.6 years.
In the time-latitude diagram, which is the experimental basis for our analysis of magnetic field variations, ripples with different polarities appear very clearly. However, upon closer inspection, it turns out that this phenomenon is quite complex, and over the course of several cycles, the parameters of variations undergo significant changes. 
	
\begin{figure}[h]

\centerline{\includegraphics[width=0.75\textwidth,clip=]{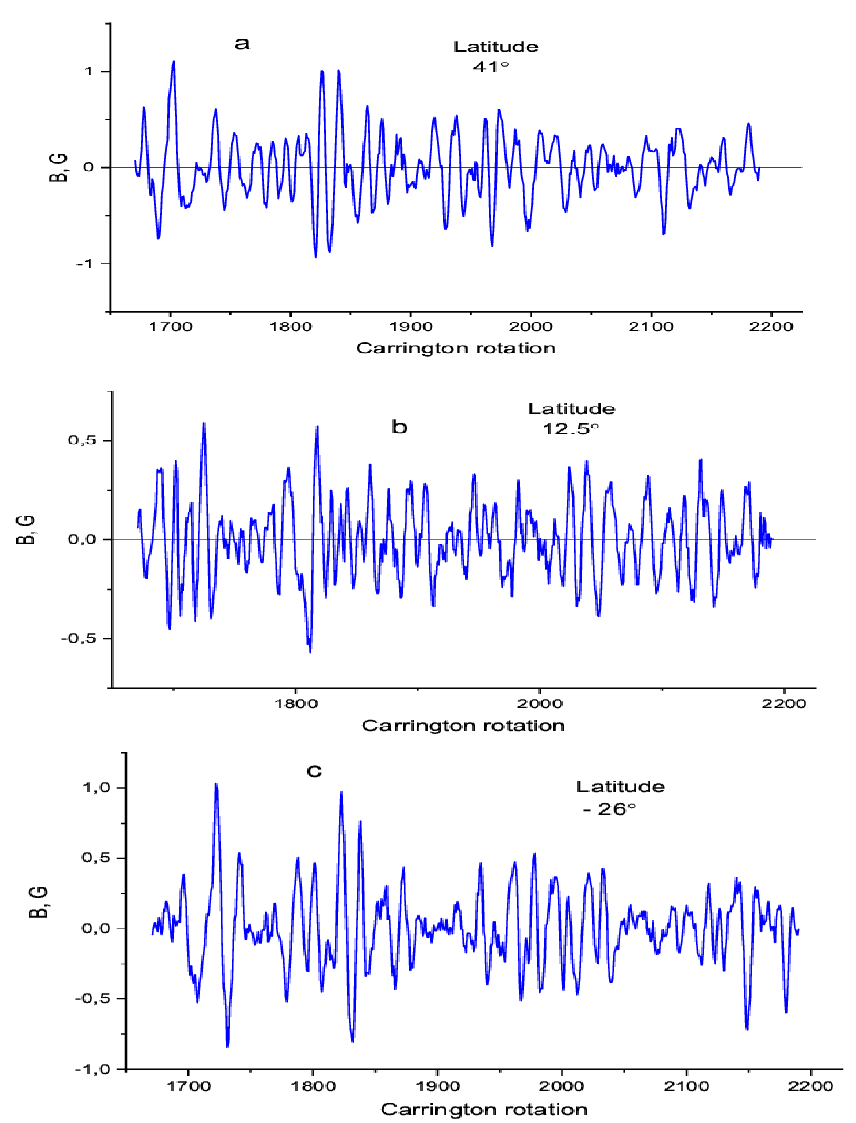}

}
  \caption{Change of magnetic field $| B |\leq 5$\,G (same as in Figure~\ref{method}c) for 4 solar cycles 21--24. Latitudes: (a) $41^\circ$,  (b) $12.5^\circ$, (c) $-26^\circ$.
}

 \label{fourcycles}                                        
   \end{figure}

For example, the lifetime of these variations (the region of the time-latitude diagram in which the variation is constantly present) varies during 1978--2016 from 6 to 11 years in the northern and southern hemispheres. Over four solar cycles, three time intervals can be distinguished, during which ripples were observed.  These intervals situated  between two RTTPs include the decline, minimum and rise of solar activity: 1982--1989, 1992--1998, 2001--2011. Thus, at each latitude we have 3 intervals (regions) that we will analyze. To make it more convenient to compare the northern and southern hemispheres, 
we denote by N1, N2, N3 these three intervals for the latitudes of the northern hemisphere and by S1, S2, S3 the three latitude
intervals of the southern hemisphere.
In order to obtain quantitative estimates of the period and amplitude of the ripples  it was necessary to remove the variations with the shortest and longest periods from the original data, since it is clear that ripples are associated with variations having a period from 0.5 years to several years.

To highlight these variations, we use the following procedure:  we determine the long period trend by averaging the magnetic field values over 21 rotations
(Figure~\ref{method}a) then we calculate the deviation from the trend (Figure~\ref{method}b). After subtraction of the trend resulting values were averaged over 5 Carrington rotations 
(Figure~\ref{method}c).

This technique was used in \citep{vern2} where the periodic structure of ripples for one fixed latitude ($-33^\circ$) was investigated. In this work, we apply the same technique for latitudes: $\pm 41^\circ$, $\pm 33^\circ$, $\pm 26^\circ$,  $\pm 19^\circ$,  $\pm 12.5^\circ$, $\pm 6^\circ$, $-0.3^\circ$ (equator).
Figure~\ref{fourcycles} for latitudes $41^\circ$, $12.5^\circ$, and $-26^\circ$ shows  magnetic field changes over 4 solar cycles. You can see signs of an 11-year cycle in the amplitude of the variations. This cyclicity is noticeable for latitudes $41^\circ$, and $\sim -26^\circ$ where it manifests itself as a decrease in the amplitude of variation during low SA years. For the latitude of $12.5^\circ$ the amplitude of variations is high in years of low activity.
	
\begin{figure}[h]

\centerline{\includegraphics[width=0.75\textwidth,clip=]{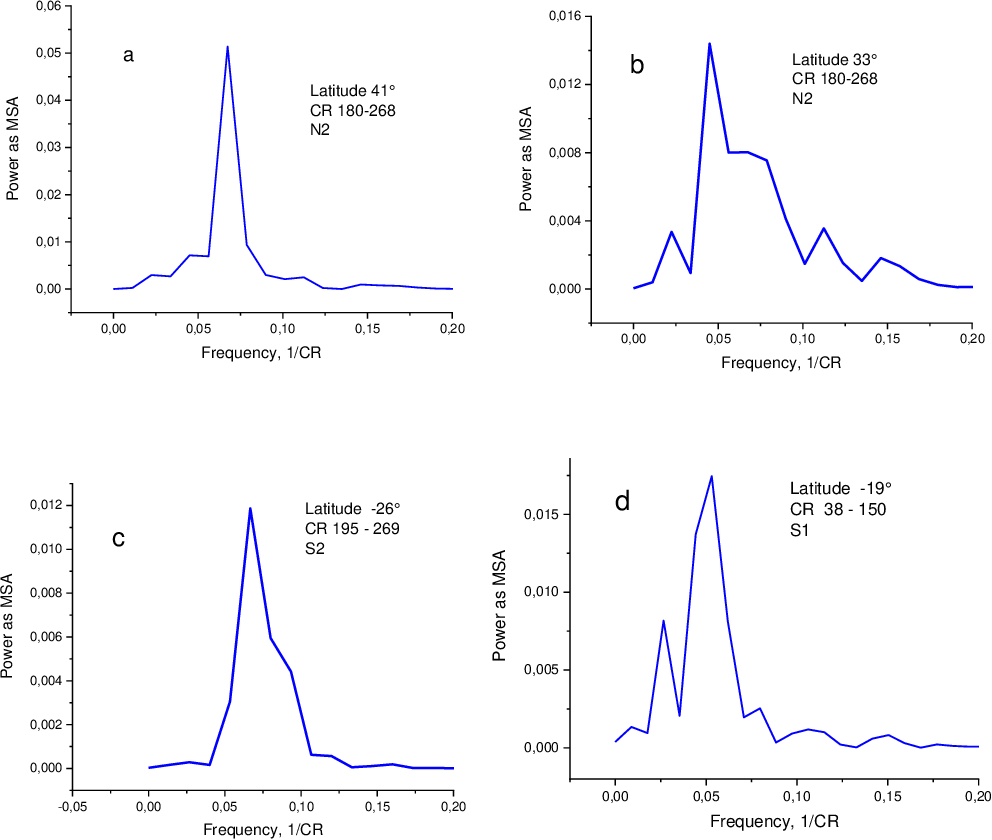}

}
\caption{Examples of Fast Fourier Transforms for latitudes/intervals: (a) $41^\circ$/N2; (b) $33^\circ$/N2; (c) $-26^\circ$/S2; (d) $-19^\circ$/S1.
}

 \label{Fourier}                                        
   \end{figure}

Along with  the 11-year cycle, short-period variations are constantly present, which appear even after this simple data processing (trend subtraction and smoothing by 5 Carrington rotations): one can see that the fields change sign cyclically.
To analyze the frequency spectrum of ripples, we used the Origin computational package. The data set after processing by the FFT program from this package contains a large number of different parameters that characterize the features of the temporal changes of the investigated phenomenon. One of the widely used characteristics of spectral analysis is the power spectrum represented in the Origin package by the function ``power spectrum as mean square amplitude (MSA)''. Each data series consisting of magnetic field values   allocated at one of the selected latitudes was processed by the program. As a result we obtained about 40 frequency spectra (for 6 regions N1, N2, N3 and S1, S2, S3 and 13 different latitudes). Examples of Fourier spectra are shown in Figures~\ref{Fourier}a--d. It should be noted that sharp amplitude peaks accounted for a relatively small fraction of the spectra: about 1/3 of the total number of spectra.
The processing of data on the distribution of the magnetic field using the Fourier transform made it possible to establish a number of properties of ripples. A histogram of the ripples period  
(Figure~\ref{Period}) showed that the variation of periods lies in the range from 0.5 to 4 years with a maximum at 1.2 years. The ripples group with a period of more than 2.5 years, apparently, should be attributed to statistical outliers. After excluding this group (8 cases out of 72), the average period would be $T = 1.30 \pm 0.06$ years.
\begin{figure}[h]

\centerline{\includegraphics[width=0.75\textwidth,clip=]{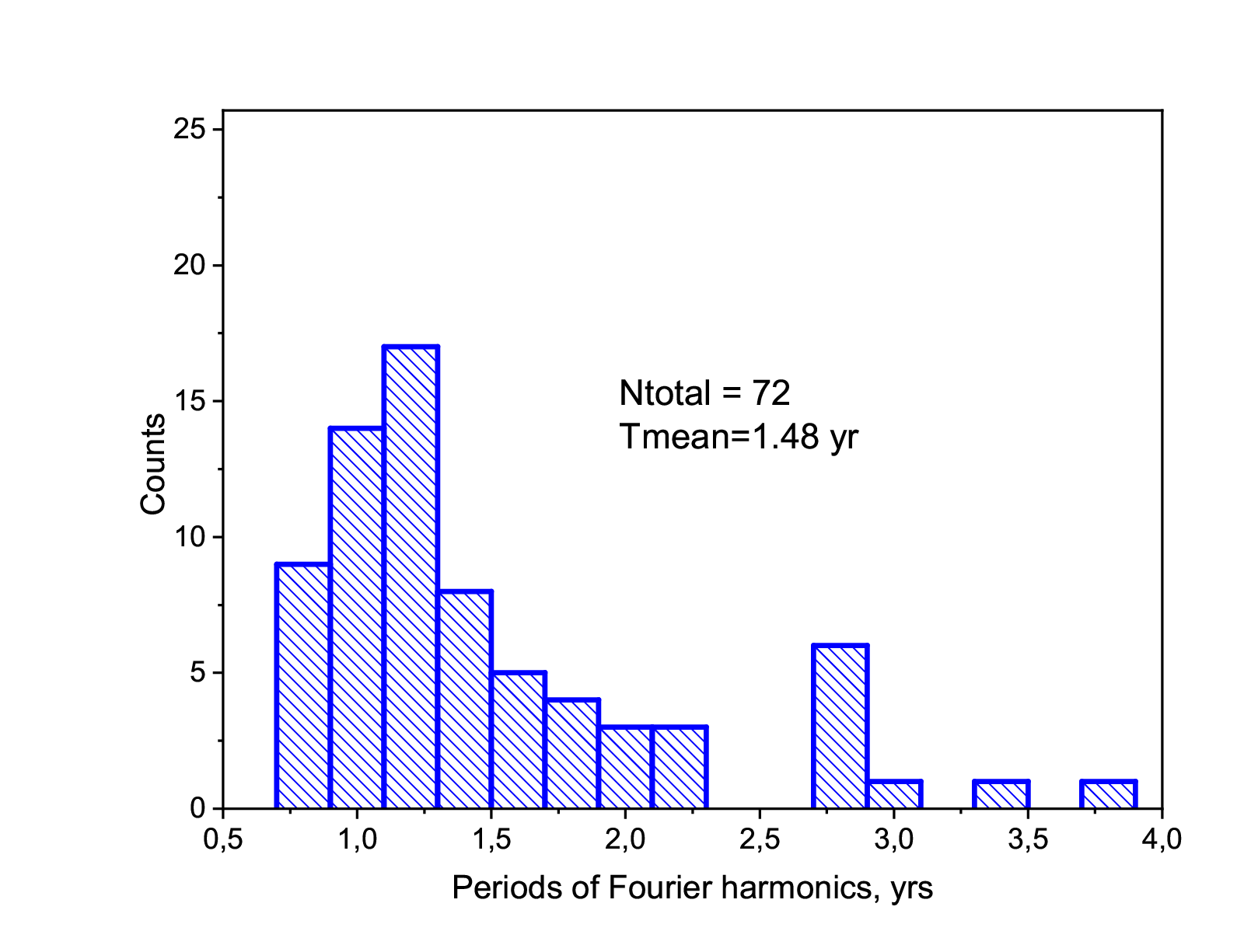}

}
\caption{Histogram of the periods of harmonics obtained by the Fast Fourier Transform for all the latitudes considered. The maximum of the histogram falls on 1.2 years.
}
						
 \label{Period}                                        
   \end{figure}

The space-time intervals we have chosen relate to different hemispheres and different time periods. The lifetime of ripples falls on the part of the solar cycle enclosed between two maxima, i.e., the period of constant sign of the polar field. Figure~\ref{amp} shows a histogram of amplitudes for each of the N or S intervals (indicated by arrows at the top of the histogram). The period denoted by N1 corresponds to the polar field sign ``$-$'' while S1 corresponds to the sign ``+''. So for each of the 6 intervals, the polarity of the polar field during this period is indicated 
(Figure~\ref{amp}). It turned out that the amplitude of the variations is connected with the sign of the polar field during the corresponding interval. Figure~\ref{amp} shows that the amplitude of ripples is significantly higher for those intervals in which the polar field had a positive sign. The sum of all amplitude values   for intervals with the polar field sign ``+'' is 1.05, while for intervals with the polar field sign ``$-$'' this sum was 0.35, i.e., three times less. This effect confirms the connection of ripples variations with the polarity of the 22-year magnetic cycle.

\begin{figure}[h]

\centerline{\includegraphics[width=0.75\textwidth,clip=]{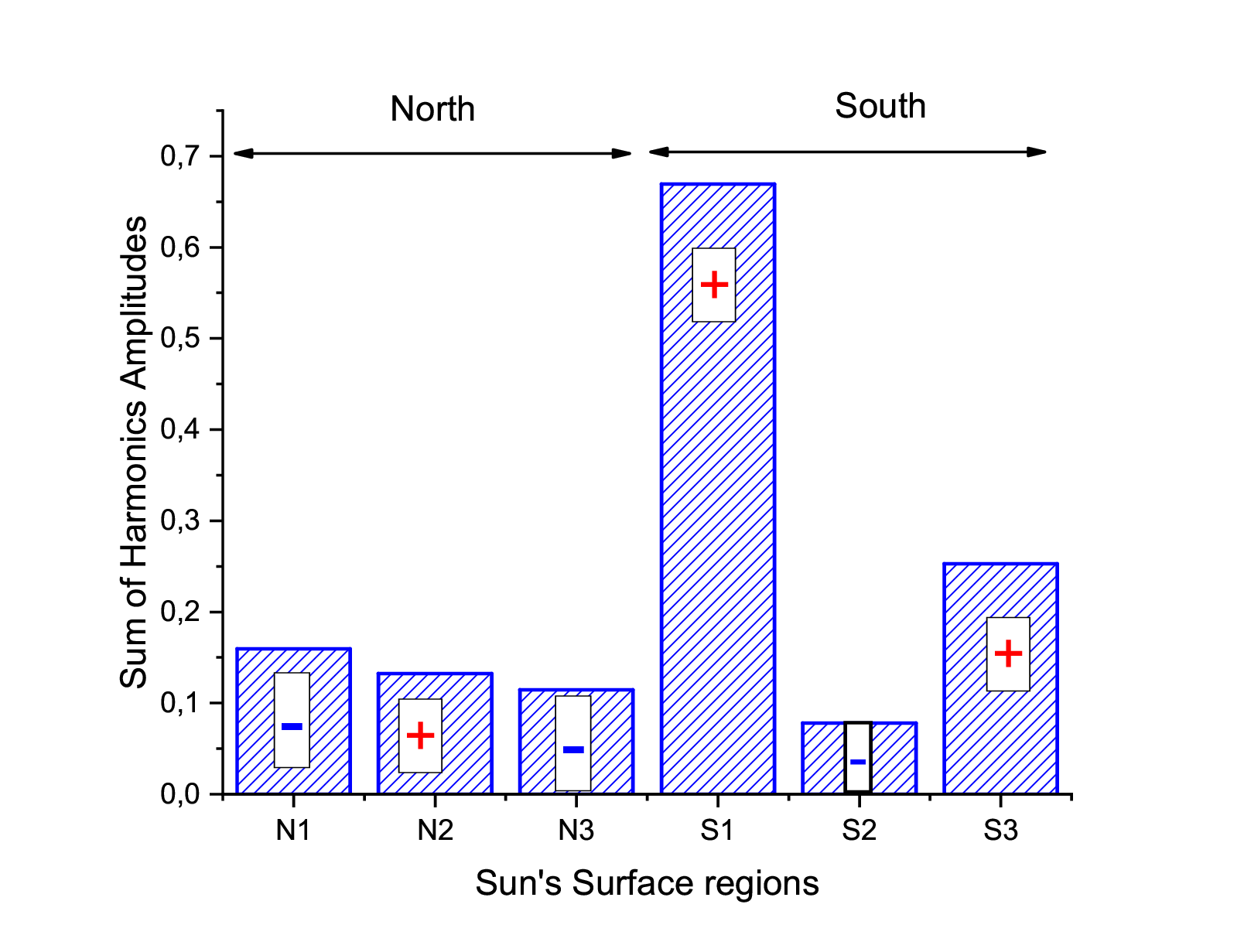}

}
  \caption{Amplitudes of harmonics obtained by fast Fourier transform for intervals N1, N2, N3 (northern hemisphere) and S1, S2, S3 (southern hemisphere).  The polar field sign is shown for each of the intervals.
}
						
 \label{amp}                                        
   \end{figure}

\section{Conclusions}
Synoptic maps of the Sun's magnetic field were used to study the properties of the ripples phenomenon (magnetic field fluxes of alternating polarity in the photosphere). By smoothing the primary data, we eliminated both the shortest variations and long-period ones from the primary data, as a result, variations of weak fields became more pronounced. In the time-latitude diagram: 6 regions with distinct cyclic structure were observed during 4 solar cycles (21--24). The processing of these data using the Fast Fourier Transform made it possible to establish a number of properties of ripples. 

A histogram of ripples showed that the variation of periods lies in the range from 0.5 to 4 years with a maximum of 1.2 years.
The appearance of ripples falls on the part of the solar cycle enclosed between two maxima, i.e., on the period of constant sign of the polar field. It turned out that the amplitude of the variations is associated with the sign of the polar field during the corresponding interval. The amplitude of ripples is significantly higher for those intervals in which the polar field has positive sign. The sum of all amplitude values for intervals with the polar field sign ``+'' is three times greater than for intervals with the polar field sign ``$-$''. This effect confirms the connection of ripples variations with the polarity of the 22-year magnetic cycle.

\section{Acknowledgements}
The NSO/Kitt Peak data used here are produced cooperatively by NSF/NOAO,NASA/GSFC, and NO-AA/SEL (ftp://nispdata.nso.edu/kpvt
/synoptic/mag/). Data acquired by SOLIS instruments were operated by NISP/NSO/AURA/NSF. https://magmap.nso.edu/solis/archive.html


\begin{thebibliography}{}
\bibitem[\protect\citeauthoryear{Altrock}{2014}]{alt}
Altrock, R.C.: 2014, {\it Solar Phys.} {\bf 289}, 623.

\bibitem[\protect\citeauthoryear{Getachew, Virtanen, and Mursula}{2019a}]{geta1}
Getachew T., Virtanen I.,  Mursula K: {2019a}, {\it Astrophys. J.} {\bf 874}, 116.

\bibitem[\protect\citeauthoryear{Getachew, Virtanen, and Mursula}{2019b}]{geta2}
Getachew T., Virtanen I.,  Mursula K: {2019b}, {\it Geophys. Res. Lett. } {\bf 46}, 9327.

\bibitem[\protect\citeauthoryear{Gopalswamy \etal}{2016}]{gopa}
Gopalswamy, N., Yashiro, S.,  Akiyama, S.: 2016, {\it Astrophys. J. Letters}, {\bf823}, L15.

\bibitem[\protect\citeauthoryear {Harvey}{1996}]{harv}
Harvey, J.: 1996, {\it http://www.noao.edu/noao/staff/jharvey/pole.ps}.

\bibitem[\protect\citeauthoryear{Mordvinov \etal}{2016}]{mord}
Mordvinov, A.V., Pevtsov, A.A., Bertello, L.,  Petrie, G.J.D.: 2016, {\it Solar-Terrestrial Physics}. {\bf  2}, Iss. 1, 3.

\bibitem[\protect\citeauthoryear{Mursula, Getachew, and Virtanen}{2021}]{murs}
Mursula, K.,  Getachew, T., Virtanen, I.: 2021, {\it Astron. Astrophys.} {\bf  645}, id. A47.

\bibitem[\protect\citeauthoryear{Petrie}{2015}]{petr}
Petrie, G.J.D.: 2015, {\it Liv. Rev. Sol. Phys}. {\bf 12}, 5.

\bibitem[\protect\citeauthoryear{Ulrich and Tran}{ 2013}]{ulri}
Ulrich, R.K., Tran, T.: 2013, {\it Astrophys. J.} {\bf 768}, 189.

\bibitem[\protect\citeauthoryear{Vecchio \etal}{2012}]{vecc}
Vecchio,  A., Laurenza, M., Meduri, D., Carbone, V., Storini, M.: 2012, {\it Astrophys. J.}. {\bf 749}, 27.

\bibitem[\protect\citeauthoryear{Vernova, Tyasto, and Baranov}{2022}]{vern1}
Vernova, E.S., Tyasto, M.I., Baranov, D.G.: 2022, {\it Geomagnetism and Aeronomy}. {\bf  62(7)},  945.

\bibitem[\protect\citeauthoryear{Vernova, Tyasto, and Baranov}{2023}]{vern2}
Vernova, E.S., Tyasto, M.I., Baranov, D.G.: 2023, {\it  Solar Phys.}. {\bf 298}, id 69. 


\end{thebibliography}
\end{document}